\documentstyle[amssymb,preprint, aps]{revtex}

\begin{document}
\draft
\title{Theory of Open Quantum Systems as Applied to Spin Relaxation in Solids}
\author{Vadim I. Puller, Lev G. Mourokh and Norman J.M. Horing}
\address{Department of Physics and Engineering Physics, \\
Stevens Institute of Technology, Hoboken, NJ 07030 USA}
\author{Anatoly Yu. Smirnov}
\address{D-Wave Systems Inc., 320-1985 W. Broadway \\
Vancouver, British Columbia Canada V6J 4Y3}
\date{\today}
\maketitle

\begin{abstract}
{We employ the method of the theory of open quantum systems to analyze spin
relaxation and decoherence in semiconductors in the presence of a magnetic
field. We derive a set of Bloch equations for electron spin with a fully
microscopic determination of longitudinal and transverse relaxation times.
Electron scattering from optical and acoustic phonons and random impurities
is taken into account. We obtain explicit expressions for the spin
relaxation times in terms of material constants and coupling strengths,
exhibiting formal agreement with earlier treatments in the zero magnetic
field limit with microscopic specification of their phenomenological
parameters.}
\end{abstract}


\narrowtext

\section{Introduction}

Modern statistical physics encompasses a wide range of scientific problems
from elementary particles to the evolution of the Universe. Its methods
facilitate the determination of the macroscopic characteristics of complex
nonequilibrium systems without having to exactly analyze the highly
complicated individual microscopic dynamics of each constituent part. In
many cases the system of interest can be separated into two distinct
subsystems. One of them, which we will term the ''{\it dynamical subsystem}%
'', has only few degrees of freedom, whereas the other, the ''{\it bath}'',
has , in an ideal case, the infinite number of them. The bath is a source of
random perturbations (fluctuations) for the dynamical subsystem, also
functioning as an absorber of dissipative energy. This separation is at the
core of the theory of open quantum systems (TOQS) based on papers of
Schwinger \cite{Schwinger}, developed in Refs. \cite{EK,ES,EMS} and
successfully applied to radiation damping \cite{RD} and some problems in
solid state physics \cite{SS}.

In the present paper we demonstrate the applicability of TOQS to the problem
of spin dynamics in semiconductors, which has been extensively studied over
the last decade both experimentally \cite{ExpBulk} and theoretically \cite
{Teor} in conjunction with numerous proposals for a variety of spin-based
devices \cite{Awsh}. The most promising materials for device purposes, the
III-V and II-VI compounds, have been shown to have spin relaxation rates
dominated by the D'yakonov-Perel' (DP) mechanism \cite{DP} at moderate
temperatures and low hole concentrations. In contrast to earlier treatments,
we develop a fully microscopic stochastic theory of spin dynamics in the
presence of an external magnetic field, taking account of pertinent
scattering mechanisms. Our analysis is based on an innovative implementation
of the two step relaxation process corresponding to the relaxation time
hierarchy involved in (a) electron thermalization, and (b) spin relaxation.
In the first stage of solution, we determine the relaxation rates and
fluctuation characteristics of electron orbital motion due to coupling to
optical and acoustic phonons and random impurities as a bath. Spin
relaxation dynamics (the slowest process in the system) can be neglected in
stage (a). The second stage, (b), proceeds with the analysis of the spin
relaxation process due to spin-orbit interaction, wherein the orbital
degrees of freedom are considered as an ''effective heat bath'', having the
characteristics determined in stage (a). The orbital motion can be
considered as an intermediary transmitting fluctuations from the
phonon/impurity bath to electron spin, and, also, transferring the
dissipated energy flow in the opposite direction. A set of Bloch equations
with two distinct relaxation times (longitudinal relaxation time, $T_{1}$,
responsible for spin magnetic moment relaxation, and transverse relaxation
time, $T_{2}$, responsible for decoherence) is derived in this second stage.
In both stages of our analysis we employ the TOQS method.

The structure of the paper is as follows. In Section II we present a brief
outline of TOQS and the principal relations used in subsequent Sections. The
microscopic formulation of the spin relaxation problem is given in Section
III. In Section IV we analyze the orbital electron dynamics in the presence
of an external magnetic field with coupling to optic and acoustic phonons
and random impurities. The spin-orbit interaction is examined in Section V,
where we derive the Bloch equations on a microscopic basis. A summary of
this work is given in Section VI.

\section{Theory of Open Quantum Systems}

The full Hamiltonian of the system, separated into a ''dynamical subsystem''
and a ''bath'', can be written as 
\begin{equation}
H=H_{0}+H_{B}+H_{int},  \label{H}
\end{equation}
where $H_{0}$ is the Hamiltonian of the dynamical subsystem, $H_{B}$ is the
Hamiltonian of the bath and $H_{int}$ describes their interaction. We can
write the last term in a product form 
\begin{equation}
H_{int}=-\sum_{\alpha }F_{\alpha }(t)Q_{\alpha }(t),  \label{Hint}
\end{equation}
where $F_{\alpha }(t)$ is, in the general case, a nonlinear function of the
dynamical subsystem variables, $Q_{\alpha }(t)$ is a function of the bath
variables, and the summation index $\alpha $ can refer to either coordinate
projections or a mode index. One can see from $H_{int}$ that both the
dynamical subsystem and the bath influence each other. Moreover, $F_{\alpha
} $ plays a role of a generalized force conjugate to the generalized bath
coordinate, $Q_{\alpha }$ and vice versa (Figure1).

In the case of Gaussian statistics of the unperturbed bath variables, or
when the coupling of the dynamical subsystem to the bath is weak, the full
Heisenberg operator of the bath is given by \cite{ES} 
\begin{equation}
Q_{\alpha }^{h}(t)=Q_{\alpha }^{0}(t)+\int dt_{1}\varphi _{\alpha \beta
}(t,t_{1})F_{\beta }(t_{1}),  \label{Response}
\end{equation}
where 
\begin{equation}
\varphi _{\alpha \beta }(t,t_{1})=\langle {\frac{i}{\hbar }}[Q_{\alpha
}^{0}(t),Q_{\beta }^{0}(t_{1})]_{-}\rangle \eta (t-t_{1})  \label{Fi}
\end{equation}
is a linear response function, or the retarded Green's function of the
unperturbed bath variables. $\eta \left( t-t_{1}\right) $ is the Heaviside
unit step function. Another important function is the correlation function
of the unperturbed bath variables 
\begin{equation}
M_{\alpha \beta }(t,t_{1})=\langle {\frac{1}{2}}[Q_{\alpha }^{0}(t),Q_{\beta
}^{0}(t_{1})]_{+}\rangle =M_{\alpha \beta }(\tau ).  \label{eM}
\end{equation}
Here and below we use the notation $\left[ ...,...\right] _{-}$ for a
commutator, and $\left[ ...,...\right] _{+}$ for an anticommutator. The
linear response function of Eq.(\ref{Fi}), and the correlation function of
Eq.(\ref{eM}), are related by means of the fluctuation-dissipation theorem 
\begin{equation}
S_{\alpha \beta }(\omega )=\hbar \chi _{\alpha \beta }^{\prime \prime
}(\omega )coth{\frac{\hbar \omega }{2k_{B}T},}  \label{FDT}
\end{equation}
where 
\begin{equation}
S_{\alpha \beta }(\omega )=\int d\tau e^{-i\omega \tau }M_{\alpha \beta
}(\tau ),
\end{equation}
and 
\begin{equation}
\chi _{\alpha \beta }(\omega )=\int d\tau e^{-i\omega \tau }\varphi _{\alpha
\beta }(\tau )=\chi _{\alpha \beta }^{\prime }(\omega )+i\chi _{\alpha \beta
}^{\prime \prime }(\omega ).
\end{equation}
To derive the fluctuation-dissipation theorem we employed the relation 
\begin{equation}
\langle \lbrack Q_{\beta }^{0}(0),Q_{\alpha }^{0}(\omega )]_{+}\rangle
=\langle \lbrack Q_{\beta }^{0}(0),Q_{\alpha }^{0}(\omega )]_{-}\rangle coth{%
\frac{\hbar \omega }{2k_{B}T},}  \label{coth}
\end{equation}
based on the Gibbs distribution and, therefore, the fluctuation-dissipation
theorem, in the present form, is valid only for systems in equilibrium.

The Heisenberg equation of motion for an arbitrary operator of the dynamical
subsystem, $A(t)$,is given by 
\begin{equation}
\dot{A}(t)={\frac{1}{i\hbar }}[A(t),H_{0}]_{-}-{\frac{1}{i\hbar }}%
[A(t),F_{\alpha }(t)]_{-}Q_{\alpha }^{h}(t),
\end{equation}
where $Q_{\alpha }^{h}(t)$ is a Heisenberg bath operator including the
influence of the dynamical subsystem. Substituting the expression of Eq. (%
\ref{Response}) for this operator, and taking account of the fact that the
operator of the dynamical subsystem is commutative with $Q_{\alpha }^{h}(t)$
at the same moment of time (only), we obtain 
\begin{equation}
\dot{A}(t)={\frac{1}{i\hbar }}[A(t),H_{0}]_{-}-{\frac{1}{2}}[Q_{\alpha
}^{0}(t),Y_{\alpha }(t)]_{+}-{\frac{1}{2}}\int dt_{1}\varphi _{\alpha \beta
}(t,t_{1})[Y_{\alpha }(t),F_{\beta }(t_{1})]_{+},
\end{equation}
where 
\[
Y_{\alpha }(t)={\frac{1}{i\hbar }}[A(t),F_{\alpha }(t)]_{-}. 
\]
Considering $Y_{\alpha }(t)$ as a function of the unperturbed bath variables
and employing the quantum analog of the Furutsu-Novikov theorem \cite{ES},
we find that 
\begin{equation}
\langle {\frac{1}{2}}[Q_{\alpha }^{0}(t),Y_{\alpha }(t)]_{+}\rangle =\int
dt_{1}M_{\alpha \beta }(t,t_{1})\langle {\frac{\delta Y_{\alpha }(t)}{\delta
Q_{\beta }^{0}(t_{1})}}\rangle ,
\end{equation}
and eliminate the appearance of the bath variables using functional
derivatives as described by Efremov and Smirnov \cite{ES} 
\begin{equation}
\langle {\frac{\delta Y_{\alpha }(t)}{\delta Q_{\beta }^{0}(t_{1})}}\rangle
=\langle {\frac{i}{\hbar }}[Y_{\alpha }(t),F_{\beta }(t_{1})]_{-}\rangle
\eta (t-t_{1}).
\end{equation}
Introducing the fluctuation source, $\xi (t)$, 
\begin{equation}
\xi (t)=-{\frac{1}{2}}[Q_{\alpha }^{0}(t),Y_{\alpha }(t)]_{+}+\int_{-\infty
}^{t}dt_{1}M_{\alpha \beta }(t,t_{1}){\frac{i}{\hbar }}[Y_{\alpha
}(t),F_{\beta }(t_{1})]_{-},
\end{equation}
with zero mean value $(\langle \xi (t)\rangle =0)$, we find that the
equation for the arbitrary operator $A(t)$ of the dynamical subsystem has
the form 
\begin{eqnarray}
\dot{A}(t) &=&{\frac{1}{i\hbar }}[A(t),H_{0}]_{-}-{\frac{1}{2}}\int_{-\infty
}^{\infty }dt_{1}\varphi _{\alpha \beta }(t,t_{1})[Y_{\alpha }(t),F_{\beta
}(t_{1})]_{+}-  \nonumber \\
&&-\int_{-\infty }^{t}dt_{1}M_{\alpha \beta }(t,t_{1}){\frac{i}{\hbar }}%
[Y_{\alpha }(t),F_{\beta }(t_{1})]_{-}+\xi (t).
\end{eqnarray}
In the case of weak coupling between the dynamical subsystem and the bath,
the correlator of the fluctuation force is given by 
\begin{equation}
\left\langle \frac{1}{2}\left[ \xi (t),\xi (t_{1})\right] _{+}\right\rangle
=M_{\alpha \beta }(t,t_{1})\left\langle {\frac{1}{2}}[Y_{\alpha
}(t),Y_{\beta }(t_{1})]_{+}\right\rangle +R_{\alpha \beta
}(t,t_{1})\left\langle {\frac{1}{2}}[Y_{\alpha }(t),Y_{\beta
}(t_{1})]_{-}\right\rangle ,
\end{equation}
where 
\begin{equation}
R_{\alpha \beta }(t,t_{1})=\langle {\frac{1}{2}}[Q_{\alpha }^{0}(t),Q_{\beta
}^{0}(t_{1})]_{-}\rangle =\frac{\hbar }{2i}\left( \varphi _{\alpha \beta
}(t,t_{1})-\varphi _{\beta \alpha }(t_{1},t)\right) .
\end{equation}

\section{Formulation of spin dynamics}

We start from the model Hamiltonian 
\begin{equation}
\widehat{H}=H_{orbital}+H_{spin}+U_{e-ph}+U_{e-i}+U_{DP}+H_{ph},  \label{Ham}
\end{equation}
describing an electron with spin in the presence of a magnetic field
directed along the $z$-axis, where the magnetic field and its vector
potential are given by 
\begin{equation}
{\bf B}=(0,0,B),\text{ }{\bf A}=(-\frac{By}{2};\frac{Bx}{2};0).
\end{equation}
The first term in the Hamiltonian (\ref{Ham}) is responsible for kinetic
electron orbital motion, 
\begin{equation}
H_{orbital}=\frac{mV_{x}^{2}}{2}+\frac{mV_{y}^{2}}{2}+\frac{mV_{z}^{2}}{2},
\end{equation}
where the velocity component operators of the electron in a magnetic field \
can be written as 
\begin{eqnarray}
V_{x} &=&\frac{1}{m}\left( p_{x}-\frac{m\omega _{c}y}{2}\right) ,\text{ }%
V_{y}=\frac{1}{m}\left( p_{y}+\frac{m\omega _{c}x}{2}\right) ,\text{ }V_{z}=%
\frac{p_{z}}{m}, \\
\left[ V_{x},V_{y}\right] _{-} &=&-\frac{i\hbar \omega _{c}}{m}.  \nonumber
\end{eqnarray}
The next term, describing spin motion in the presence of a magnetic field
(Zeeman term), is given by 
\begin{equation}
H_{spin}=\frac{1}{2}g\mu _{B}\left( \overrightarrow{{\bf \sigma }}\cdot {\bf %
B}\right) =\frac{\hbar \omega _{B}}{2}\sigma _{z},
\end{equation}
where $g$ is the crystal $g$-factor (in particular, $g=-0.44$ for GaAs), $%
\mu _{B}=\left| e\right| \hbar /2m_{0}c$ is the Bohr magneton ($m_{0}$ is
the mass of a free electron, the effective mass, $m$, is $0.067m_{0}$ for
GaAs), $\omega _{B}=g\mu _{B}B/\hbar $ is the frequency of the spin
precession induced by the magnetic field, and $\overrightarrow{{\bf \sigma }}%
=\left( \sigma _{x},\sigma _{y},\sigma _{z}\right) $ is the standard set of
Pauli matrices.

The free phonon Hamiltonian is given by 
\begin{equation}
H_{B}=\sum_{{\bf k}}\hbar \omega _{{\bf k}}(b_{{\bf k}}^{+}b_{{\bf k}}+{%
\frac{1}{2}}),
\end{equation}
where $\hbar \omega _{{\bf k}}$ is the phonon energy and $b_{{\bf k}}^{+}$
and $b_{{\bf k}}$ are the creation and annihilation operators of a phonon in
the ${\bf k}$-mode, respectively. The interaction of an electron with random
impurities is described by the potential 
\begin{equation}
U_{e-i}({\bf r})=-\frac{1}{L^{3/2}}\sum_{{\bf k}}U_{{\bf k}}e^{i{\bf kr}},
\end{equation}
where $U_{{\bf k}}$ are the spatial Fourier components of the impurity
potential, and $L^{3}$ denotes the volume of the crystal. The
electron-phonon interaction is given by 
\begin{equation}
U_{e-ph}({\bf r},t)=-\frac{1}{L^{3/2}}\sum_{{\bf k}}i\zeta (b_{{\bf k}%
}(t)-b_{-{\bf k}}^{+}(t))e^{i{\bf kr}},
\end{equation}
where $\zeta $ presents the strength of the electron-phonon coupling. As
mentioned above, the principal mechanism of spin-orbit interaction is the
D'yakonov-Perel' term, $U_{DP}$. In semiconductor crystals lacking inversion
symmetry, the effective mass Hamiltonian includes terms cubic in electron
quasimomentum \cite{DP}. The presence of these terms induces a random
magnetic field ${\bf \Omega }$ given by 
\begin{equation}
{\bf \Omega }=\frac{\alpha \overrightarrow{{\bf \varkappa }}}{\hbar m^{3/2}%
\sqrt{2E_{g}}},
\end{equation}
where 
\begin{equation}
\varkappa _{x}=m^{3}V_{x}\left( V_{y}^{2}-V_{z}^{2}\right) ,\text{ }%
\varkappa _{y}=m^{3}V_{y}\left( V_{z}^{2}-V_{x}^{2}\right) ,\text{ }%
\varkappa _{z}=m^{3}V_{z}\left( V_{x}^{2}-V_{y}^{2}\right) ,
\end{equation}
$E_{g}$ is the energy gap, and $\alpha $ is a coefficient representing the
strength of the spin-orbit coupling ($\alpha \approx 0.07$ in GaAs). The
random magnetic field induced by the orbital motion interacts with electron
spin through the interaction Hamiltonian 
\begin{equation}
U_{DP}=\frac{\hbar }{2}\left( \overrightarrow{{\bf \sigma }}{\bf \cdot
\Omega }\right)
\end{equation}

It is evident from the Hamiltonian (\ref{Ham}) that there is no {\it direct }%
coupling of electron spin degrees of freedom to phonons and impurities.
Therefore, to analyze spin relaxation dynamics, we consider the orbital
degrees of freedom as an intermediary that (a) transfers fluctuations from
the phonon/impurity bath to electron spin, and (b) transfers energy from
spin degrees of freedom to heat bath. This two-step process is illustrated
in Figure 2. Correspondingly, we address the analysis in two stages, and in
the first stage we determine the statistical characteristics of orbital
motion dictated by the phonon/impurity bath, neglecting the presence of the
electron spin. The pertinent first-stage Hamiltonian, $H^{\left( I\right) }$%
, is given by 
\begin{equation}
H^{\left( I\right) }=H_{orbital}+H_{int}^{\left( I\right) }+H_{ph},
\label{1st}
\end{equation}
where the first-stage interaction Hamiltonian, $H_{int}^{\left( I\right) }$,
has the form 
\begin{equation}
H_{int}^{\left( I\right) }=U_{e-ph}+U_{e-i}=-L^{-3/2}\sum_{{\bf k}}\left(
i\zeta (b_{{\bf k}}(t)-b_{-{\bf k}}^{+}(t))+U_{{\bf k}}\right) e^{i{\bf kr}}.
\end{equation}
Comparing this expression to Eq. (\ref{Hint}), we take the summation index $%
\alpha $ of Eq. (\ref{Hint}), as the mode number, ${\bf k}$, and the
function $L^{-3/2}e^{i{\bf kr}}$ plays the role of \ $F_{\alpha }(t)$ \ of
Eq. (\ref{Hint}), whereas the bath variable,\ $Q_{\alpha }(t)$, is $\left(
i\zeta (b_{{\bf k}}(t)-b_{-{\bf k}}^{+}(t))+U_{{\bf k}}\right) $.

In the second stage of analysis we employ orbital motion as an ''effective
heat bath'' having the characteristics determined in the first stage. The
second-stage Hamiltonian, $H^{\left( II\right) }$, can be written as 
\begin{equation}
H^{\left( II\right) }=H_{spin}+H_{int}^{\left( II\right) }+H_{bath},
\label{2nd}
\end{equation}
where $H_{int}^{\left( II\right) }=U_{DP}$ and $H_{bath}$\ may be taken as
the remainder of the Hamiltonian (\ref{Ham}). To compare $U_{DP}$ to Eq. (%
\ref{Hint}), we rewrite $U_{DP}$ as

\begin{equation}
U_{DP}=-\sigma _{x}Q_{x}(t)-\sigma _{y}Q_{y}(t)-\sigma _{z}Q_{z}(t),
\label{2ndint}
\end{equation}
where the ''effective heat bath'' variables are given by 
\begin{eqnarray}
Q_{x}(t) &=&-\frac{\alpha m^{3/2}}{2\sqrt{2\varepsilon _{g}}}V_{x}(t)\left(
V_{y}^{2}(t)-V_{z}^{2}(t)\right) ,\text{ }  \label{Q} \\
Q_{y}(t) &=&-\frac{\alpha m^{3/2}}{2\sqrt{2\varepsilon _{g}}}V_{y}(t)\left(
V_{z}^{2}(t)-V_{x}^{2}(t)\right) ,  \nonumber \\
Q_{z}(t) &=&-\frac{\alpha m^{3/2}}{2\sqrt{2\varepsilon _{g}}}V_{z}(t)\left(
V_{x}^{2}(t)-V_{y}^{2}(t)\right) .  \nonumber
\end{eqnarray}
In this, there is summation over coordinate projections and the ''dynamic
subsystem'' variables are the electron spin projections.

\section{Stage 1: Orbital dynamics.}

In this step we employ TOQS to determine the statistical characteristics of
electron orbital motion described by the operator equations \cite{SS} 
\begin{eqnarray}
\left( \frac{d}{dt}+\gamma _{0}\right) V_{x}(t)+\left( \omega _{c}+\delta
\right) V_{y}(t) &=&\xi _{x}(t),  \label{Velocity} \\
\left( \frac{d}{dt}+\gamma _{0}\right) V_{y}(t)-\left( \omega _{c}-\delta
\right) V_{x}(t) &=&\xi _{y}(t),  \nonumber
\end{eqnarray}
and 
\[
\left( \frac{d}{dt}+\gamma _{z}\right) V_{z}(t)=\xi _{z}(t), 
\]
where $V_{x}(t),V_{y}(t),V_{z}(t)$ are electron velocity operator
components, and $\omega _{c}=\left| e\right| B/mc$ is the cyclotron
frequency. The electron-bath interaction determines the relaxation rates, $%
\gamma _{0}$,$\gamma _{z}$, the frequency shift, $\delta $, and the
fluctuation sources, $\xi _{x}(t),\xi _{y}(t),\xi _{z}(t)$, involved in Eq. (%
\ref{Velocity}), \cite{SS}. The Fourier transforms of the velocity
correlation functions are given by 
\begin{equation}
\left\langle \frac{1}{2}\left[ V_{x}(\omega );V_{x}\right] _{+}\right\rangle
=\left\langle \frac{1}{2}\left[ V_{y}(\omega );V_{y}\right]
_{+}\right\rangle =\frac{K_{\bot }(\omega )}{2}\left( \frac{1}{\left( \omega
-\omega _{c}\right) ^{2}+\gamma _{0}^{2}}+\frac{1}{\left( \omega +\omega
_{c}\right) ^{2}+\gamma _{0}^{2}}\right) ,  \label{xx}
\end{equation}
\begin{equation}
\left\langle \frac{1}{2}\left[ V_{x}(\omega );V_{y}\right] _{+}\right\rangle
=-\left\langle \frac{1}{2}\left[ V_{y}(\omega );V_{x}\right]
_{+}\right\rangle =\frac{K_{\bot }(\omega )}{2i}\left( \frac{1}{\left(
\omega -\omega _{c}\right) ^{2}+\gamma _{0}^{2}}-\frac{1}{\left( \omega
+\omega _{c}\right) ^{2}+\gamma _{0}^{2}}\right) ,  \label{xy}
\end{equation}
and 
\begin{equation}
\left\langle \frac{1}{2}\left[ V_{z}(\omega );V_{z}\right] _{+}\right\rangle
=\frac{K_{z}(\omega )}{\omega ^{2}+\gamma _{z}^{2}},  \label{zz}
\end{equation}
where 
\begin{equation}
K_{\bot }(\omega )=\int d\left( t-t_{1}\right) e^{i\omega \left(
t-t_{1}\right) }\left\langle \frac{1}{2}\left[ \xi _{x}(t),\xi _{x}(t_{1})%
\right] _{+}\right\rangle =\int d\left( t-t_{1}\right) e^{i\omega \left(
t-t_{1}\right) }\left\langle \frac{1}{2}\left[ \xi _{y}(t),\xi _{y}(t_{1})%
\right] _{+}\right\rangle ,
\end{equation}
and 
\begin{equation}
K_{z}(\omega )=\int d\left( t-t_{1}\right) e^{i\omega \left( t-t_{1}\right)
}\left\langle \frac{1}{2}\left[ \xi _{z}(t),\xi _{z}(t_{1})\right]
_{+}\right\rangle .
\end{equation}

To be specific we take account of the contributions of polar optical
phonons, deformational acoustic phonons and random impurities to electron
orbital dynamics. In a first approximation in their (weak) coupling
strengths to orbital motion, these contributions can be treated as mutually
independent and the overall values of the damping rates and the time-Fourier
transforms of the correlation functions are given by (superscripts $OP$ and $%
AP$\ denote the contributions of optical and acoustic phonons, respectively,
and $I$ denotes that of random impurities): 
\begin{eqnarray}
\gamma _{0} &=&\gamma _{0}^{OP}+\gamma _{0}^{I}+\gamma _{0}^{AP}, \\
\gamma _{z} &=&\gamma _{z}^{OP}+\gamma _{z}^{I}+\gamma _{z}^{AP},  \nonumber
\\
K_{\bot }(\omega ) &=&K_{\bot }^{OP}(\omega )+K_{\bot }^{I}(\omega )+K_{\bot
}^{AP}(\omega ),  \nonumber \\
K_{z}(\omega ) &=&K_{z}^{OP}(\omega )+K_{z}^{I}(\omega )+K_{z}^{AP}(\omega ),
\nonumber
\end{eqnarray}
The microscopic expressions of all these quantities are determined using the
method of Ref. \cite{SS}, and are given below:

The response and correlation functions for polar optical phonons are 
\begin{equation}
\varphi _{{\bf k}}^{OP}(\tau )=\frac{4\pi \Omega _{0}e^{2}}{k^{2}\epsilon
^{\ast }}\sin \left( \Omega _{0}\tau \right) \eta (\tau ),\text{ }M_{{\bf k}%
}^{OP}(\tau )=\frac{\hbar }{2}\frac{4\pi \Omega _{0}e^{2}}{k^{2}\epsilon
^{\ast }}\cos \left( \Omega _{0}\tau \right) \coth (\frac{\hbar \Omega _{0}}{%
2k_{B}T}),
\end{equation}
resulting in 
\[
\gamma _{0}^{OP}=\frac{1}{2\sqrt{2\pi }}\frac{\Omega _{0}e^{2}}{m\epsilon
^{\ast }}\int_{0}^{+\infty }dk_{\perp }k_{\perp }^{3}\int_{0}^{+\infty
}dk_{z}\frac{\tau _{c}^{3}(k_{\perp },k_{z})}{k^{2}}{\Bbb G}^{OP}\left(
k_{\bot },k_{z}\right) 
\]
and 
\[
\gamma _{z}^{OP}=\frac{1}{\sqrt{2\pi }}\frac{\Omega _{0}e^{2}}{m\epsilon
^{\ast }}\int_{0}^{+\infty }dk_{\perp }k_{\perp }\int_{0}^{+\infty }dk_{z}%
\frac{\tau _{c}^{3}(k_{\perp },k_{z})}{k^{2}}k_{z}^{2}{\Bbb G}^{OP}\left(
k_{\bot },k_{z}\right) , 
\]
where 
\begin{eqnarray}
{\Bbb G}^{OP}\left( k_{\bot },k_{z}\right) &=&\left\{ \left( \coth (\frac{%
\hbar \Omega _{0}}{2k_{B}T})+1\right) \left( \omega _{k}+\Omega _{0}\right)
\exp \left[ -\frac{1}{2}\left( \omega _{k}+\Omega _{0}\right) ^{2}\tau
_{c}^{2}(k_{\perp },k_{z})\right] +\right. \\
&&\left. +\left( \coth (\frac{\hbar \Omega _{0}}{2k_{B}T})-1\right) \left(
\omega _{k}-\Omega _{0}\right) \exp \left[ -\frac{1}{2}\left( \omega
_{k}-\Omega _{0}\right) ^{2}\tau _{c}^{2}(k_{\perp },k_{z})\right] \right\} .
\nonumber
\end{eqnarray}
Furthermore, we obtain 
\begin{equation}
K_{\bot }^{OP}(\omega )=\frac{1}{4\sqrt{2\pi }}\frac{\hbar \Omega _{0}e^{2}}{%
m^{2}\epsilon ^{\ast }}\int_{0}^{+\infty }dk_{\perp }k_{\perp
}^{3}\int_{0}^{+\infty }dk_{z}\frac{\tau _{c}(k_{\perp },k_{z})}{k^{2}}{\Bbb %
K}^{OP}\left( k_{\perp },k_{z},\omega \right)
\end{equation}
and 
\begin{equation}
K_{z}^{OP}(\omega )=\frac{1}{2\sqrt{2\pi }}\frac{\hbar \Omega _{0}e^{2}}{%
m^{2}\epsilon ^{\ast }}\int_{0}^{+\infty }dk_{\perp }k_{\perp
}\int_{0}^{+\infty }dk_{z}\frac{\tau _{c}(k_{\perp },k_{z})}{k^{2}}k_{z}^{2}%
{\Bbb K}^{OP}\left( k_{\perp },k_{z},\omega \right) ,
\end{equation}
where 
\begin{eqnarray}
{\Bbb K}^{OP}\left( k_{\perp },k_{z},\omega \right) &=&\left\{ \left( \coth (%
\frac{\hbar \Omega _{0}}{2k_{B}T})+1\right) \left( \exp \left[ -\frac{1}{2}%
\left( \omega +\omega _{k}+\Omega _{0}\right) ^{2}\tau _{c}^{2}(k_{\perp
},k_{z})\right] +\right. \right. \\
&&\left. +\exp \left[ -\frac{1}{2}\left( \omega -\omega _{k}-\Omega
_{0}\right) ^{2}\tau _{c}^{2}(k_{\perp },k_{z})\right] \right) +  \nonumber
\\
&&+\left( \coth (\frac{\hbar \Omega _{0}}{2k_{B}T})-1\right) \left( \exp %
\left[ -\frac{1}{2}\left( \omega +\omega _{k}-\Omega _{0}\right) ^{2}\tau
_{c}^{2}(k_{\perp },k_{z})\right] +\right.  \nonumber \\
&&\left. \left. +\exp \left[ -\frac{1}{2}\left( \omega -\omega _{k}+\Omega
_{0}\right) ^{2}\tau _{c}^{2}(k_{\perp },k_{z})\right] \right) \right\} , 
\nonumber
\end{eqnarray}
$\Omega _{0}$ is the optical phonon frequency, $1/\epsilon ^{\ast
}=1/\epsilon _{\infty }-1/\epsilon _{0}$, ($\epsilon _{\infty }$ and $%
\epsilon _{0}$ are the hf and static permittivities of the crystal,
respectively).

The response and correlation functions of the acoustic phonons are given by 
\begin{equation}
\varphi _{{\bf k}}^{AP}(\tau )=\frac{D^{2}k}{\rho u}\sin \left( uk\tau
\right) \eta (\tau ),\text{ }M_{{\bf k}}^{AP}(\tau )=\frac{\hbar }{2}\frac{%
D^{2}k}{\rho u}\cos \left( uk\tau \right) \coth (\frac{\hbar uk}{2k_{B}T}),
\end{equation}
which provide the contribution to the damping rate as 
\begin{equation}
\gamma _{0}^{AP}=\frac{1}{8\pi \sqrt{2\pi }}\frac{D^{2}}{m\rho u}%
\int_{0}^{+\infty }dk_{\perp }k_{\perp }^{3}\int_{0}^{+\infty }dk_{z}k\tau
_{c}^{3}(k_{\perp },k_{z}){\Bbb G}^{AP}\left( k_{\bot },k_{z}\right)
\end{equation}
and 
\begin{equation}
\gamma _{z}^{AP}=\frac{1}{4\pi \sqrt{2\pi }}\frac{D^{2}}{m\rho u}%
\int_{0}^{+\infty }dk_{\perp }k_{\perp }\int_{0}^{+\infty
}dk_{z}kk_{z}^{2}\tau _{c}^{3}(k_{\perp },k_{z}){\Bbb G}^{AP}\left( k_{\bot
},k_{z}\right) ,
\end{equation}
where 
\begin{eqnarray}
{\Bbb G}^{AP}\left( k_{\bot },k_{z}\right) &=&\left\{ \left( \coth (\frac{%
\hbar uk}{2k_{B}T})+1\right) \left( \omega _{k}+uk\right) \exp \left[ -\frac{%
1}{2}\left( \omega _{k}+uk\right) ^{2}\tau _{c}^{2}(k_{\perp },k_{z})\right]
+\right. \\
&&\left. +\left( \coth (\frac{\hbar uk}{2k_{B}T})-1\right) \left( \omega
_{k}-uk\right) \exp \left[ -\frac{1}{2}\left( \omega _{k}-uk\right) ^{2}\tau
_{c}^{2}(k_{\perp },k_{z})\right] \right\} .  \nonumber
\end{eqnarray}
Accordingly, 
\begin{equation}
K_{\bot }^{AP}(\omega )=\frac{1}{16\pi \sqrt{2\pi }}\frac{\hbar D^{2}}{%
m^{2}\rho u}\int_{0}^{+\infty }dk_{\perp }k_{\perp }^{3}\int_{0}^{+\infty
}dk_{z}k\tau _{c}(k_{\perp },k_{z}){\Bbb K}^{AP}\left( k_{\perp
},k_{z},\omega \right) ,
\end{equation}
and 
\begin{equation}
K_{z}^{AP}(\omega )=\frac{1}{8\pi \sqrt{2\pi }}\frac{\hbar D^{2}}{m^{2}\rho u%
}\int_{0}^{+\infty }dk_{\perp }k_{\perp }\int_{0}^{+\infty
}dk_{z}kk_{z}^{2}\tau _{c}(k_{\perp },k_{z}){\Bbb K}^{AP}\left( k_{\perp
},k_{z},\omega \right) ,
\end{equation}
where 
\begin{eqnarray}
{\Bbb K}^{AP}\left( k_{\perp },k_{z},\omega \right) &=&\left\{ \left( \coth (%
\frac{\hbar uk}{2k_{B}T})+1\right) \left( \exp \left[ -\frac{1}{2}\left(
\omega +\omega _{k}+uk\right) ^{2}\tau _{c}^{2}(k_{\perp },k_{z})\right]
+\right. \right. \\
&&\left. +\exp \left[ -\frac{1}{2}\left( \omega -\omega _{k}-uk\right)
^{2}\tau _{c}^{2}(k_{\perp },k_{z})\right] \right) +  \nonumber \\
&&+\left( \coth (\frac{\hbar uk}{2k_{B}T})-1\right) \left( \exp \left[ -%
\frac{1}{2}\left( \omega +\omega _{k}-uk\right) ^{2}\tau _{c}^{2}(k_{\perp
},k_{z})\right] +\right.  \nonumber \\
&&\left. \left. +\exp \left[ -\frac{1}{2}\left( \omega -\omega
_{k}+uk\right) ^{2}\tau _{c}^{2}(k_{\perp },k_{z})\right] \right) \right\} .
\nonumber
\end{eqnarray}
Here, $D$ is the deformation potential , $\rho $ is the crystal density, and 
$u$ is the sound velocity.

For the case of static random impurities, there is not any response
function, due to the lack of dynamics. The correlation function is given by 
\begin{equation}
M_{{\bf k}}^{I}=\frac{2e^{4}n_{t}^{\ast }}{\pi \epsilon _{0}^{2}\left(
k^{2}+r_{0}^{-2}\right) ^{2}},
\end{equation}
where $r_{0}$ is the screening radius, $n_{t}^{\ast }=\sum_{\alpha
}n_{\alpha }Z_{a}^{2}$, $n_{\alpha }$ is the impurity concentration for
species $\alpha $, and $Z_{\alpha }$ is their charge number. The relaxation
rates associated with impurities have the forms 
\begin{equation}
\gamma _{0}^{I}=\frac{1}{2\pi ^{2}\sqrt{2\pi }}\frac{e^{4}n_{t}^{\ast }}{%
m^{2}\epsilon _{0}^{2}}\int_{0}^{+\infty }dk_{\perp }k_{\perp
}^{3}\int_{0}^{+\infty }dk_{z}\frac{k^{2}\tau _{c}^{3}(k_{\perp },k_{z})}{%
\left( k^{2}+r_{0}^{-2}\right) ^{2}}\exp \left( -\frac{\omega _{k}^{2}\tau
_{c}^{2}(k_{\perp },k_{z})}{2}\right) ,
\end{equation}
and 
\begin{equation}
\gamma _{z}^{I}=\frac{1}{\pi ^{2}\sqrt{2\pi }}\frac{e^{4}n_{t}^{\ast }}{%
m^{2}\epsilon _{0}^{2}}\int_{0}^{+\infty }dk_{\perp }k_{\perp
}\int_{0}^{+\infty }dk_{z}\frac{k_{z}^{2}k^{2}\tau _{c}^{3}(k_{\perp },k_{z})%
}{\left( k^{2}+r_{0}^{-2}\right) ^{2}}\exp \left( -\frac{\omega _{k}^{2}\tau
_{c}^{2}(k_{\perp },k_{z})}{2}\right) .
\end{equation}
Finally, the contributions to the Fourier transforms of the fluctuation
force \ correlators can be written as 
\begin{eqnarray}
K_{\bot }^{I}(\omega ) &=&\frac{1}{2\pi ^{2}\sqrt{2\pi }}\frac{%
e^{4}n_{t}^{\ast }}{m^{2}\epsilon _{0}^{2}}\int_{0}^{+\infty }dk_{\perp
}k_{\perp }^{3}\int_{0}^{+\infty }dk_{z}\frac{\tau _{c}(k_{\perp },k_{z})}{%
\left( k^{2}+r_{0}^{-2}\right) ^{2}}\cdot \\
&&\cdot \left( \exp \left[ -\frac{1}{2}\left( \omega +\omega _{k}\right)
^{2}\tau _{c}^{2}(k_{\perp },k_{z})\right] +\exp \left[ -\frac{1}{2}\left(
\omega -\omega _{k}\right) ^{2}\tau _{c}^{2}(k_{\perp },k_{z})\right]
\right) ,  \nonumber
\end{eqnarray}
and 
\begin{eqnarray}
K_{z}^{I}(\omega ) &=&\frac{1}{\pi ^{2}\sqrt{2\pi }}\frac{e^{4}n_{t}^{\ast }%
}{m^{2}\epsilon _{0}^{2}}\int_{0}^{+\infty }dk_{\perp }k_{\perp
}\int_{0}^{+\infty }dk_{z}\frac{k_{z}^{2}\tau _{c}(k_{\perp },k_{z})}{\left(
k^{2}+r_{0}^{-2}\right) ^{2}}\cdot \\
&&\cdot \left( \exp \left[ -\frac{1}{2}\left( \omega +\omega _{k}\right)
^{2}\tau _{c}^{2}(k_{\perp },k_{z})\right] +\exp \left[ -\frac{1}{2}\left(
\omega -\omega _{k}\right) ^{2}\tau _{c}^{2}(k_{\perp },k_{z})\right]
\right) ,  \nonumber
\end{eqnarray}

In all these formulae we have used the notation $k=\sqrt{k_{\perp
}^{2}+k_{z}^{2}},$ $\omega _{k}=\hbar k^{2}/2m,$ $\tau _{c}^{-2}(k_{\perp
},k_{z})=k_{\perp }^{2}\left\langle V_{x}^{2}\right\rangle
+k_{z}^{2}\left\langle V_{z}^{2}\right\rangle $. $\left\langle
V_{x}^{2}\right\rangle $ and $\left\langle V_{z}^{2}\right\rangle $ can be
determined self-consistently using $\left\langle V_{x}^{2}\right\rangle
=K_{\bot }(\omega _{c})/2\gamma _{0}$ and $\left\langle
V_{z}^{2}\right\rangle =K_{z}(0)/2\gamma _{z}$ \cite{SS}. All these results
are obtained for relatively weak magnetic fields, $\omega _{c}\tau _{c}<<1$.
However, the energy shift, $\delta $, due to electron-bath coupling is even
smaller, $\delta <<\omega _{c}$, and will be neglected.

\section{Stage 2: Spin dynamics.}

In this section we examine the spin relaxation process in the presence of
the ''effective bath'' with variables given by Eq. (\ref{Q}). Their response
functions and correlation functions have the form ($i,j=x,y,z$) 
\begin{equation}
M_{ij}\left( t,t_{1}\right) =\langle {\frac{1}{2}}%
[Q_{i}(t),Q_{j}(t_{1})]_{+}\rangle 
\end{equation}
\begin{equation}
\varphi _{ij}\left( t,t_{1}\right) =\langle {\frac{i}{\hbar }}%
[Q_{i}(t),Q_{j}(t_{1})]_{-}\rangle \eta (t-t_{1})
\end{equation}
It is evident that these functions are of the sixth order of the electron
velocity projections. Using the quantum analog of the Furutsu-Novikov
theorem for Gaussian variables \cite{ES}, they can be represented in terms
of pair correlators. The resulting expressions for the spectral functions
are given by 
\begin{equation}
S_{ij}\left( \omega \right) =\int d\left( t-t_{1}\right) e^{i\omega \left(
t-t_{1}\right) }M_{ij}\left( t,t_{1}\right) ,
\end{equation}
\begin{eqnarray}
S_{xx}(\omega ) &=&S_{yy}(\omega )=\lambda ^{2}\left\langle \frac{1}{2}\left[
V_{x}(\omega ),V_{x}\right] _{+}\right\rangle \left[ \left( \left\langle
V_{z}^{2}\right\rangle -\left\langle V_{x}^{2}\right\rangle \right)
^{2}+4\left\langle V_{x}V_{y}\right\rangle \left\langle
V_{x}V_{y}\right\rangle \right] +  \label{Sxx} \\
&&+\lambda ^{2}\int \frac{d\omega _{1}}{2\pi }\int \frac{d\omega _{2}}{2\pi }%
\Xi \left( \omega _{1},\omega _{2},\omega -\omega _{1}-\omega _{2}\right)
\cdot   \nonumber \\
&&\cdot \left\{ 2\left\langle \frac{1}{2}\left[ V_{x}(\omega _{1}),V_{x}%
\right] _{+}\right\rangle \left\langle \frac{1}{2}\left[ V_{x}(\omega
_{2}),V_{x}\right] _{+}\right\rangle \left\langle \frac{1}{2}\left[
V_{x}(\omega -\omega _{1}-\omega _{2}),V_{x}\right] _{+}\right\rangle
+\right.   \nonumber \\
&&2\left\langle \frac{1}{2}\left[ V_{x}(\omega _{1}),V_{x}\right]
_{+}\right\rangle \left\langle \frac{1}{2}\left[ V_{z}(\omega _{2}),V_{z}%
\right] _{+}\right\rangle \left\langle \frac{1}{2}\left[ V_{z}(\omega
-\omega _{1}-\omega _{2}),V_{z}\right] _{+}\right\rangle -  \nonumber \\
&&\left. -4\left\langle \frac{1}{2}\left[ V_{x}(\omega _{1}),V_{x}\right]
_{+}\right\rangle \left\langle \frac{1}{2}\left[ V_{x}(\omega _{2}),V_{y}%
\right] _{+}\right\rangle \left\langle \frac{1}{2}\left[ V_{x}(\omega
-\omega _{1}-\omega _{2}),V_{y}\right] _{+}\right\rangle \right\} , 
\nonumber
\end{eqnarray}
\begin{eqnarray}
S_{xy}(\omega ) &=&-S_{yx}(\omega )=\lambda ^{2}\left\langle \frac{1}{2}%
\left[ V_{y}(\omega ),V_{x}\right] _{+}\right\rangle \left[ \left(
\left\langle V_{z}^{2}\right\rangle -\left\langle V_{x}^{2}\right\rangle
\right) ^{2}+4\left\langle V_{x}V_{y}\right\rangle \left\langle
V_{x}V_{y}\right\rangle \right] +  \label{Sxy} \\
&&+\lambda ^{2}\int \frac{d\omega _{1}}{2\pi }\int \frac{d\omega _{2}}{2\pi }%
\Xi \left( \omega _{1},\omega _{2},\omega -\omega _{1}-\omega _{2}\right)
\cdot   \nonumber \\
&&\cdot \left\{ 2\left\langle \frac{1}{2}\left[ V_{y}(\omega _{1}),V_{x}%
\right] _{+}\right\rangle \left\langle \frac{1}{2}\left[ V_{y}(\omega
_{2}),V_{x}\right] _{+}\right\rangle \left\langle \frac{1}{2}\left[
V_{y}(\omega -\omega _{1}-\omega _{2}),V_{x}\right] _{+}\right\rangle
+\right.   \nonumber \\
&&+2\left\langle \frac{1}{2}\left[ V_{y}(\omega _{1}),V_{x}\right]
_{+}\right\rangle \left\langle \frac{1}{2}\left[ V_{z}(\omega _{2}),V_{z}%
\right] _{+}\right\rangle \left\langle \frac{1}{2}\left[ V_{z}(\omega
-\omega _{1}-\omega _{2}),V_{z}\right] _{+}\right\rangle -  \nonumber \\
&&\left. -4\left\langle \frac{1}{2}\left[ V_{y}(\omega _{1}),V_{x}\right]
_{+}\right\rangle \left\langle \frac{1}{2}\left[ V_{x}(\omega _{2}),V_{x}%
\right] _{+}\right\rangle \left\langle \frac{1}{2}\left[ V_{x}(\omega
-\omega _{1}-\omega _{2}),V_{x}\right] _{+}\right\rangle \right\} , 
\nonumber
\end{eqnarray}
and 
\begin{eqnarray}
S_{zz}(\omega ) &=&\lambda ^{2}\int \frac{d\omega _{1}}{2\pi }\int \frac{%
d\omega _{2}}{2\pi }\Xi \left( \omega _{1},\omega _{2},\omega -\omega
_{1}-\omega _{2}\right) \cdot   \label{Szz} \\
&&\cdot 4\left\langle \frac{1}{2}\left[ V_{z}(\omega _{1}),V_{z}\right]
_{+}\right\rangle \left\{ \left\langle \frac{1}{2}\left[ V_{x}(\omega
_{2}),V_{x}\right] _{+}\right\rangle \left\langle \frac{1}{2}\left[
V_{x}(\omega -\omega _{1}-\omega _{2}),V_{x}\right] _{+}\right\rangle
-\right.   \nonumber \\
&&\left. -\left\langle \frac{1}{2}\left[ V_{x}(\omega _{2}),V_{y}\right]
_{+}\right\rangle \left\langle \frac{1}{2}\left[ V_{x}(\omega -\omega
_{1}-\omega _{2}),V_{y}\right] _{+}\right\rangle \right\} ,  \nonumber
\end{eqnarray}
where we defined an auxiliary function 
\begin{eqnarray}
\Xi \left( \omega _{1},\omega _{2},\omega _{3}\right)  &=&1+\tanh \left( 
\frac{\hbar \omega _{1}}{2T}\right) \tanh \left( \frac{\hbar \omega _{2}}{2T}%
\right) +  \nonumber \\
&&+\tanh \left( \frac{\hbar \omega _{1}}{2T}\right) \tanh \left( \frac{\hbar
\omega _{3}}{2T}\right) +\tanh \left( \frac{\hbar \omega _{2}}{2T}\right)
\tanh \left( \frac{\hbar \omega _{3}}{2T}\right) ,
\end{eqnarray}
which appears due to the replacement of the velocity commutators by their
anticommutators in accordance with Eq. (\ref{coth}). In Eqs. (\ref{Sxx}-\ref
{Szz}) we use the shortened notation $\left\langle V_{j}^{2}\right\rangle
=\left\langle \frac{1}{2}\left[ V_{j}(t),V_{j}(t)\right] _{+}\right\rangle
=\int \frac{d\omega }{2\pi }\left\langle \frac{1}{2}\left[ V_{j}(\omega
),V_{j}\right] _{+}\right\rangle ,$ and $\left\langle
V_{x}V_{y}\right\rangle =\left\langle \frac{1}{2}\left[ V_{x}(t),V_{y}(t)%
\right] _{+}\right\rangle =\int \frac{d\omega }{2\pi }\left\langle \frac{1}{2%
}\left[ V_{x}(\omega ),V_{y}\right] _{+}\right\rangle $.Employing a second
application of the TOQS method, we obtain equations for the average spin
projections as follows: 
\begin{eqnarray}
\frac{d}{dt}\left\langle \sigma _{x}(t)\right\rangle  &=&-\Gamma
_{xx}\left\langle \sigma _{x}(t)\right\rangle -\left( \omega _{B}+\Gamma
_{xy}\right) \left\langle \sigma _{y}(t)\right\rangle +\Gamma
_{xz}\left\langle \sigma _{x}(t)\right\rangle +\Gamma _{x}^{0}, \\
\frac{d}{dt}\left\langle \sigma _{y}(t)\right\rangle  &=&\left( \omega
_{B}+\Gamma _{yx}\right) \left\langle \sigma _{x}(t)\right\rangle -\Gamma
_{yy}\left\langle \sigma _{y}(t)\right\rangle +\Gamma _{yz}\left\langle
\sigma _{z}(t)\right\rangle +\Gamma _{y}^{0},  \nonumber \\
\frac{d}{dt}\left\langle \sigma _{z}(t)\right\rangle  &=&\Gamma
_{zx}\left\langle \sigma _{x}(t)\right\rangle +\Gamma _{zy}\left\langle
\sigma _{y}(t)\right\rangle -\Gamma _{zz}\left\langle \sigma
_{z}(t)\right\rangle +\Gamma _{z}^{0},  \nonumber
\end{eqnarray}
where the coefficients are given by 
\begin{eqnarray}
\Gamma _{xx} &=&\frac{4}{\hbar ^{2}}\int_{-\infty }^{t}dt_{1}\left\{
M_{yy}\left( t,t_{1}\right) \cos \left[ \omega _{B}\left( t-t_{1}\right) %
\right] +M_{yx}\left( t,t_{1}\right) \sin \left[ \omega _{B}\left(
t-t_{1}\right) \right] +M_{zz}\left( t,t_{1}\right) \right\} , \\
\Gamma _{xy} &=&\frac{4}{\hbar ^{2}}\int_{-\infty }^{t}dt_{1}\left\{
M_{yy}\left( t,t_{1}\right) \sin \left[ \omega _{B}\left( t-t_{1}\right) %
\right] -M_{yx}\left( t,t_{1}\right) \cos \left[ \omega _{B}\left(
t-t_{1}\right) \right] \right\} ,  \nonumber \\
\Gamma _{xz} &=&\frac{4}{\hbar ^{2}}\int_{-\infty }^{t}dt_{1}\left\{
M_{zx}\left( t,t_{1}\right) \cos \left[ \omega _{B}\left( t-t_{1}\right) %
\right] -M_{zy}\left( t,t_{1}\right) \sin \left[ \omega _{B}\left(
t-t_{1}\right) \right] \right\} ,  \nonumber \\
\Gamma _{x}^{0} &=&\frac{2}{\hbar }\int_{-\infty }^{t}dt_{1}\left\{ \varphi
_{zx}\left( t,t_{1}\right) \sin \left[ \omega _{B}\left( t-t_{1}\right) %
\right] +\varphi _{zy}\left( t,t_{1}\right) \cos \left[ \omega _{B}\left(
t-t_{1}\right) \right] -\varphi _{yz}\left( t,t_{1}\right) \right\} , 
\nonumber
\end{eqnarray}
\begin{eqnarray}
\Gamma _{yx} &=&\frac{4}{\hbar ^{2}}\int_{-\infty }^{t}dt_{1}\left\{
M_{xx}\left( t,t_{1}\right) \sin \left[ \omega _{B}\left( t-t_{1}\right) %
\right] +M_{xy}\left( t,t_{1}\right) \cos \left[ \omega _{B}\left(
t-t_{1}\right) \right] \right\} , \\
\Gamma _{yy} &=&\frac{4}{\hbar ^{2}}\int_{-\infty }^{t}dt_{1}\left\{
M_{xx}\left( t,t_{1}\right) \cos \left[ \omega _{B}\left( t-t_{1}\right) %
\right] -M_{xy}\left( t,t_{1}\right) \sin \left[ \omega _{B}\left(
t-t_{1}\right) \right] +M_{zz}\left( t,t_{1}\right) \right\} ,  \nonumber \\
\Gamma _{yz} &=&\frac{4}{\hbar ^{2}}\int_{-\infty }^{t}dt_{1}\left\{
M_{zx}\left( t,t_{1}\right) \sin \left[ \omega _{B}\left( t-t_{1}\right) %
\right] +M_{zy}\left( t,t_{1}\right) \cos \left[ \omega _{B}\left(
t-t_{1}\right) \right] \right\} ,  \nonumber \\
\Gamma _{y}^{0} &=&\frac{2}{\hbar }\int_{-\infty }^{t}dt_{1}\left\{ \varphi
_{zy}\left( t,t_{1}\right) \sin \left[ \omega _{B}\left( t-t_{1}\right) %
\right] -\varphi _{zx}\left( t,t_{1}\right) \cos \left[ \omega _{B}\left(
t-t_{1}\right) \right] +\varphi _{xz}\left( t,t_{1}\right) \right\} , 
\nonumber
\end{eqnarray}
and 
\begin{eqnarray}
\Gamma _{zx} &=&\frac{4}{\hbar ^{2}}\int_{-\infty }^{t}dt_{1}M_{xz}\left(
t,t_{1}\right) , \\
\Gamma _{zy} &=&\frac{4}{\hbar ^{2}}\int_{-\infty }^{t}dt_{1}M_{yz}\left(
t,t_{1}\right) ,  \nonumber \\
\Gamma _{zz} &=&\frac{4}{\hbar ^{2}}\int_{-\infty }^{t}dt_{1}\left\{ \left(
M_{yy}\left( t,t_{1}\right) +M_{xx}\left( t,t_{1}\right) \right) \cos \left[
\omega _{B}\left( t-t_{1}\right) \right] +\right.   \nonumber \\
&&\left. +\left( M_{yx}\left( t,t_{1}\right) -M_{xy}\left( t,t_{1}\right)
\right) \sin \left[ \omega _{B}\left( t-t_{1}\right) \right] \right\} , 
\nonumber \\
\Gamma _{z}^{0} &=&\frac{2}{\hbar }\int_{-\infty }^{t}dt_{1}\left\{ \left(
\varphi _{yx}\left( t,t_{1}\right) +\varphi _{xy}\left( t,t_{1}\right)
\right) \cos \left[ \omega _{B}\left( t-t_{1}\right) \right] -\right.  
\nonumber \\
&&\left. -\left( \varphi _{xx}\left( t,t_{1}\right) +\varphi _{yy}\left(
t,t_{1}\right) \right) \sin \left[ \omega _{B}\left( t-t_{1}\right) \right]
\right\} .  \nonumber
\end{eqnarray}
Only six of the above twelve coefficients, $\Gamma _{ij}$, do not vanish,
and we may identify relaxation times in terms of the nonvanishing $\Gamma
_{ij}$ as follows: 
\begin{equation}
\Gamma _{xx}=\Gamma _{yy}=\frac{1}{T_{2}},\text{ }\Gamma _{zz}=\frac{1}{T_{1}%
},\text{ }
\end{equation}
where $T_{1},T_{2}$ are the longitudinal and transverse relaxation times,
responsible for the relaxation of the magnetic moment and decoherence
respectively. Furthermore, direct calculation shows that 
\begin{equation}
\Gamma _{xy}=\Gamma _{yx}=\delta _{s},\text{ }\Gamma _{z}^{0}=\sigma
_{z}^{0}=-\tanh \left( \hbar \omega _{B}/2T\right) ,
\end{equation}
where $\delta _{s}$ may be identified as a frequency shift due to
''effective bath'' fluctuations, and $\sigma _{z}^{0}$ is the equilibrium
population difference between spin up and spin down states. With these
calculated results and identifications, the spin equations take the usual
Bloch form, 
\begin{eqnarray}
\frac{d}{dt}\left\langle \sigma _{x}(t)\right\rangle  &=&-\frac{\left\langle
\sigma _{x}(t)\right\rangle }{T_{2}}-\left( \omega _{B}+\delta _{s}\right)
\left\langle \sigma _{y}(t)\right\rangle ,  \label{Bloch} \\
\frac{d}{dt}\left\langle \sigma _{y}(t)\right\rangle  &=&\left( \omega
_{B}+\delta _{s}\right) \left\langle \sigma _{x}(t)\right\rangle -\frac{%
\left\langle \sigma _{y}(t)\right\rangle }{T_{2}},  \nonumber \\
\frac{d}{dt}\left\langle \sigma _{z}(t)\right\rangle  &=&\frac{\sigma
_{z}^{0}-\left\langle \sigma _{z}(t)\right\rangle }{T_{1}}.  \nonumber
\end{eqnarray}

In terms of the spectral functions, $S_{ij}(\omega )$, the relaxation rates
(inverse relaxation times), may be expressed as follows: 
\begin{equation}
\frac{1}{T_{1}}=\frac{4}{\hbar ^{2}}\left( S_{xx}(\omega
_{B})+iS_{xy}(\omega _{B})\right) ,  \label{T1}
\end{equation}
and 
\begin{equation}
\frac{1}{T_{2}}=\frac{2}{\hbar ^{2}}\left( S_{xx}(\omega
_{B})+iS_{xy}(\omega _{B})+S_{zz}(0)\right) .  \label{T2}
\end{equation}
Although Eqs. (\ref{T1}), (\ref{T2}) are relatively simple and yield some
interesting qualitative conclusions, a quantitative analysis of them is
difficult, because of the complexity of the expressions for the spectral
functions, $S_{ij}(\omega )$. This complexity is relieved under the
prevailing assumption of weak coupling, which permits the replacement of the
Lorenzians involved in the integrands of Eqs. (\ref{Sxx}-\ref{Szz}) by 
\begin{equation}
\Lambda (x;\varepsilon )=\frac{1}{\pi }\frac{\varepsilon }{x^{2}+\varepsilon
^{2}}\rightarrow \delta (x),\text{ when }\varepsilon \rightarrow 0,
\label{Lorenz}
\end{equation}
where $\delta (x)$ is the Dirac delta-function, and, consequently, 
\begin{eqnarray}
\int dxf(x)\Lambda (x;\varepsilon a)\Lambda (x-y;\varepsilon b) &=&\frac{1}{2%
}\Lambda (y;\varepsilon \left( a+b\right) )\left( f(y)+f(0)\right) +
\label{1} \\
&&+\frac{1}{2}\Lambda (y;\varepsilon \left( a-b\right) )\left(
f(y)-f(0)\right)  \nonumber
\end{eqnarray}
(In this, we note that in Eqs. (\ref{xx}),(\ref{xy}) and (\ref{zz}), we have 
$K_{\bot }(\omega ),K_{z}(\omega ),\gamma _{0},\gamma _{z}$ all proportional
to coupling strength $\thicksim \varepsilon $). Having employed Eqs. (\ref
{Lorenz}), (\ref{1}) in the integrals of Eqs. (\ref{Sxx}-\ref{Szz})
representing $S_{xx}(\omega ),$ $S_{xy}(\omega ),$ and $S_{zz}(\omega )$, we
obtain the results of integration as: 
\begin{eqnarray}
S_{xx}(\omega ) &=&\frac{\alpha ^{2}m^{3}}{8E_{g}}\left\langle \frac{1}{2}%
\left[ V_{x}(\omega ),V_{x}\right] _{+}\right\rangle \left[ \left(
\left\langle V_{z}^{2}\right\rangle -\left\langle V_{x}^{2}\right\rangle
\right) ^{2}+4\left\langle V_{x}V_{y}\right\rangle \left\langle
V_{x}V_{y}\right\rangle \right] +  \label{2} \\
&&+\frac{\alpha ^{2}m^{3}}{8E_{g}}\left[ R_{1}(\omega ;\omega
_{c})+Y_{1}(\omega ;\omega _{c})\right] ,  \nonumber
\end{eqnarray}
\begin{eqnarray}
S_{xy}(\omega ) &=&-\frac{\alpha ^{2}m^{3}}{8E_{g}}\left\langle \frac{1}{2}%
\left[ V_{x}(\omega ),V_{y}\right] _{+}\right\rangle \left[ \left(
\left\langle V_{z}^{2}\right\rangle -\left\langle V_{x}^{2}\right\rangle
\right) ^{2}+4\left\langle V_{x}V_{y}\right\rangle \left\langle
V_{x}V_{y}\right\rangle \right] + \\
&&+\frac{\alpha ^{2}m^{3}}{8E_{g}}\left[ R_{2}(\omega ;\omega
_{c})+Y_{2}(\omega ;\omega _{c})\right] ,  \nonumber
\end{eqnarray}
\begin{eqnarray}
S_{zz}(\omega ) &=&\frac{\alpha ^{2}m^{3}}{32E_{g}}\frac{\pi }{\gamma
_{0}^{2}\gamma _{z}}K_{\bot }(\omega _{c}) \\
&&\left\{ \Lambda (\omega -2\omega _{c};2\gamma _{0}+\gamma _{z})\left(
K_{z}(0)K_{\bot }(\omega -\omega _{c})\Xi (\omega -\omega _{c};\omega
_{c};0)+\right. \right.  \nonumber \\
&&\left. +K_{z}(\omega -2\omega _{c})K_{\bot }(\omega _{c})\Xi (\omega
-2\omega _{c};\omega _{c};\omega _{c})\right) +  \nonumber \\
&&+\Lambda (\omega -2\omega _{c};2\gamma _{0}-\gamma _{z})\left(
K_{z}(0)K_{\bot }(\omega -\omega _{c})\Xi (\omega -\omega _{c};\omega
_{c};0)-\right.  \nonumber \\
&&\left. -K_{z}(\omega -2\omega _{c})K_{\bot }(\omega _{c})Q(\omega -2\omega
_{c};\omega _{c};\omega _{c})\right) +  \nonumber \\
&&+\Lambda (\omega +2\omega _{c};2\gamma _{0}+\gamma _{z})\left(
K_{z}(0)K_{\bot }(\omega +\omega _{c})\Xi (\omega +\omega _{c};-\omega
_{c};0)+\right.  \nonumber \\
&&\left. +K_{z}(\omega +2\omega _{c})K_{\bot }(\omega _{c})\Xi (\omega
+2\omega _{c};-\omega _{c};-\omega _{c})\right) +  \nonumber \\
&&+\Lambda (\omega +2\omega _{c};2\gamma _{0}-\gamma _{z})\left(
K_{z}(0)K_{\bot }(\omega +\omega _{c})\Xi (\omega +\omega _{c};-\omega
_{c};0)-\right.  \nonumber \\
&&\left. \left. -K_{z}(\omega +2\omega _{c})K_{\bot }(\omega _{c})\Xi
(\omega +2\omega _{c};-\omega _{c};-\omega _{c})\right) \right\} ,  \nonumber
\end{eqnarray}
where we have introduced the notation 
\begin{eqnarray*}
R_{1}(\omega ,\omega _{c}) &=&\frac{\pi }{16\gamma _{0}^{3}}\left( K_{\bot
}(\omega _{c})\right) ^{2} \\
&&\left\{ 3\Lambda (\omega -3\omega _{c};3\gamma _{0})K_{\bot }(\omega
-2\omega _{c})\Xi (\omega -2\omega _{c};\omega _{c};\omega _{c})+\right. \\
&&+\Lambda (\omega -\omega _{c};3\gamma _{0})K_{\bot }(\omega -2\omega
_{c})\Xi (\omega -2\omega _{c};\omega _{c};\omega _{c})+ \\
&&+2\Lambda (\omega -\omega _{c};\gamma _{0})\left( K_{\bot }(\omega )\Xi
(\omega ;\omega _{c};-\omega _{c})-K_{\bot }(\omega -2\omega _{c})\Xi
(\omega -2\omega _{c};\omega _{c};\omega _{c})\right) + \\
&&+2\Lambda (\omega +\omega _{c};\gamma _{0})\left( K_{\bot }\left( \omega
\right) \Xi (\omega ;\omega _{c};-\omega _{c})-K_{\bot }(\omega +2\omega
_{c})\Xi (\omega +2\omega _{c};-\omega _{c};-\omega _{c})\right) + \\
&&+\Lambda (\omega +\omega _{c};3\gamma _{0})K_{\bot }(\omega +2\omega
_{c})\Xi (\omega +2\omega _{c};-\omega _{c};-\omega _{c})+ \\
&&\left. +3\Lambda (\omega +3\omega _{c};3\gamma _{0})K_{\bot }(\omega
+2\omega _{c})\Xi (\omega +2\omega _{c};-\omega _{c};-\omega _{c})\right\} ,
\end{eqnarray*}
\begin{eqnarray*}
R_{2}(\omega ,\omega _{c}) &=&-i\frac{\pi }{16\gamma _{0}^{3}}\left( K_{\bot
}(\omega _{c})\right) ^{2} \\
&&\left\{ 3\Lambda (\omega -3\omega _{c};3\gamma _{0})K_{\bot }(\omega
-2\omega _{c})\Xi (\omega -2\omega _{c};\omega _{c};\omega _{c})-\right. \\
&&-\Lambda (\omega -\omega _{c};3\gamma _{0})K_{\bot }(\omega -2\omega
_{c})\Xi (\omega -2\omega _{c};\omega _{c};\omega _{c})- \\
&&-2\Lambda (\omega -\omega _{c};\gamma _{0})\left( K_{\bot }(\omega )\Xi
(\omega ;\omega _{c};-\omega _{c})-K_{\bot }(\omega -2\omega _{c})\Xi
(\omega -2\omega _{c};\omega _{c};\omega _{c})\right) + \\
&&+2\Lambda (\omega +\omega _{c};\gamma _{0})\left( K_{\bot }(\omega )\Xi
(\omega ;\omega _{c};-\omega _{c})-K_{\bot }(\omega +2\omega _{c})\Xi
(\omega +2\omega _{c};-\omega _{c};-\omega _{c})\right) + \\
&&+\Lambda (\omega +\omega _{c};3\gamma _{0})K_{\bot }(\omega +2\omega
_{c})\Xi (\omega +2\omega _{c};-\omega _{c};-\omega _{c})- \\
&&\left. -3\Lambda (\omega +3\omega _{c};3\gamma _{0})K_{\bot }(\omega
+2\omega _{c})\Xi (\omega +2\omega _{c};-\omega _{c};-\omega _{c})\right\} ,
\end{eqnarray*}
\begin{eqnarray*}
Y_{1}(\omega ,\omega _{c}) &=&\frac{\pi }{8\gamma _{z}^{2}\gamma _{0}}%
K_{z}(0) \\
&&\left\{ \Lambda (\omega -\omega _{c};2\gamma _{z}+\gamma _{0})\left(
K_{\bot }(\omega _{c})K_{z}(\omega -\omega _{c})\Xi (\omega -\omega
_{c};\omega _{c};0)+K_{\bot }(\omega )K_{z}(0)\right) +\right. \\
&&+\Lambda (\omega -\omega _{c};2\gamma _{z}-\gamma _{0})\left( K_{\bot
}(\omega _{c})K_{z}(\omega -\omega _{c})\Xi (\omega -\omega _{c};\omega
_{c};0)-K_{\bot }(\omega )K_{z}(0)\right) + \\
&&+\Lambda (\omega +\omega _{c};2\gamma _{z}+\gamma _{0})\left( K_{\bot
}(-\omega _{c})K_{z}(\omega +\omega _{c})\Xi (\omega +\omega _{c};-\omega
_{c};0)+\widetilde{K}_{xx}(\omega )\widetilde{K}_{zz}(0)\right) + \\
&&\left. +\Lambda (\omega +\omega _{c};2\gamma _{z}-\gamma _{0})\left(
K_{\bot }(-\omega _{c})K_{z}(\omega +\omega _{c})\Xi (\omega +\omega
_{c};-\omega _{c};0)-K_{\bot }(\omega )K_{z}(0)\right) \right\} ,
\end{eqnarray*}
and 
\begin{eqnarray}
Y_{2}(\omega ,\omega _{c}) &=&\frac{i\pi }{8\gamma _{z}^{2}\gamma _{0}}%
K_{z}(0)  \label{3} \\
&&\left\{ \Lambda (\omega -\omega _{c};2\gamma _{z}+\gamma _{0})\left(
K_{\bot }(\omega _{c})K_{z}(\omega -\omega _{c})\Xi (\omega -\omega
_{c};\omega _{c};0)+K_{\bot }(\omega )K_{z}(0)\right) +\right.  \nonumber \\
&&+\Lambda (\omega -\omega _{c};2\gamma _{z}-\gamma _{0})\left( K_{\bot
}(\omega _{c})K_{z}(\omega -\omega _{c})\Xi (\omega -\omega _{c};\omega
_{c};0)-K_{\bot }(\omega )K_{z}(0)\right) -  \nonumber \\
&&-\Lambda (\omega +\omega _{c};2\gamma _{z}+\gamma _{0})\left( K_{\bot
}(-\omega _{c})K_{z}(\omega +\omega _{c})\Xi (\omega +\omega _{c};-\omega
_{c};0)+K_{\bot }(\omega )K_{z}(0)\right) -  \nonumber \\
&&\left. -\Lambda (\omega +\omega _{c};2\gamma _{z}-\gamma _{0})\left(
K_{\bot }(-\omega _{c})K_{z}(\omega +\omega _{c})\Xi (\omega +\omega
_{c};-\omega _{c};0)-K_{\bot }(\omega )K_{z}(0)\right) \right\} .  \nonumber
\end{eqnarray}
These results are very general, providing explicit expressions for the spin
relaxation times in terms of the material constants and coupling strengths.
The magnetic field has been treated as relatively weak ($\omega _{c}\tau
_{c}<<1$), but our estimates show that this range of fields includes those
used in most experiments.

To make contact with earlier theories that did not take account of the
magnetic field and used a phenomenological momentum scattering time \cite{DP}%
, we put $\omega _{c}=0$ and $\omega _{B}=0$ in Eqs. (\ref{2}-\ref{3}),
which immediately yields 
\begin{equation}
S_{xx}(\omega )=S_{xx}(\omega )=S_{zz}(\omega )=\frac{\alpha ^{2}m^{3}}{%
8E_{g}}\frac{\pi \left[ K(0)\right] ^{2}}{\gamma _{0}^{3}}K(\omega )\Lambda
(\omega ;3\alpha \gamma _{0}),
\end{equation}
where 
\begin{equation}
K_{\bot }(\omega )=K_{z}(\omega )=K(\omega ),\text{ }\gamma _{0}=\gamma _{z},
\end{equation}
and 
\begin{equation}
S_{xy}(\omega )=0.
\end{equation}
In this zero field limit the relaxation times are equal and determined by 
\begin{equation}
\frac{1}{T_{1}}=\frac{1}{T_{2}}=\frac{4}{\hbar ^{2}}S_{xx}(0).
\end{equation}
In equilibrium, the rate of velocity fluctuations is defined only by
temperature, and we obtain the result 
\begin{equation}
\left\langle V_{x}^{2}\right\rangle =\frac{K(0)}{2\gamma _{0}}=\frac{k_{B}T}{%
m},
\end{equation}
where $T$ is the Kelvin temperature. This yields the common relaxation rate
as 
\begin{equation}
\frac{1}{T_{1,2}}=\frac{4}{3}\frac{\alpha ^{2}}{\hbar ^{2}E_{g}}\frac{1}{%
\gamma _{0}}\left( k_{B}T\right) ^{3},  \label{DP}
\end{equation}
which may be brought into coincidence with the expression usually used for
the DP mechanism \cite{ExpBulk,Teor,DP} by the change of notation $q\tau
_{p}\rightarrow \frac{4}{3}\frac{1}{\gamma _{0}}$, with $\tau _{p}$ a
phenomenological momentum scattering time. It should be noted that the
choice of coefficient $q$ is unclear, with various authors assigning values
in the range $q=0.8-2.7$, with $q\approx 2.7$ for scattering by
deformational acoustic phonons and $q\approx 0.8$ for scattering by polar
optical phonons \cite{Teor,DP}. On the other hand, our microscopic analysis
provides a clearly defined value with account of all scattering mechanisms.
Moreover, we provide results for finite \ magnetic field derived on a
microscopic basis, rather than relying on the ansatz of Ref. \cite{DP},
which has $T_{1}(B)=T_{2}(B)$, 
\begin{equation}
\frac{1}{T_{1,2}(B)}=\frac{1}{T_{1,2}(B=0)}\frac{1}{1+\left( \omega _{B}\tau
_{p}\right) ^{2}}.
\end{equation}
In contradistinction to this, our analysis in the presence of a magnetic
field shows that the transverse and longitudinal spin relaxation times are
not equal (Eqs. (\ref{T1},\ref{T2})).

\section{Summary}

In Summary, we have analyzed electron spin relaxation dynamics and
decoherence in bulk semiconductors with an applied external magnetic field.
In the absence of direct coupling of spin degrees of freedom to the
phonon/impurity heat bath, we have used the orbital motion as an
intermediary, transferring fluctuations from the bath to electron spin and,
transmitting the dissipated energy from spin to the bath. To accomplish
this, the two-stage procedure of solution has been employed: In the first
stage, the fluctuation characteristics of the orbital motion were determined
and the orbital degrees of freedom were used in the second stage as an
''effective heat bath'' for spin dynamics. In both stages of the analysis,
we employed the method of the theory of open quantum systems, obtaining a
set of Bloch equations having longitudinal and transverse spin relaxation
times, $T_{1}$ and $T_{2}$, determined on a microscopic basis. Our results
provide analytic definition of the phenomenological parameters employed in
earlier zero-field theories, in terms of material constants and coupling
strengths. Moreover, this analysis of electron spin dynamics in finite
magnetic field yields explicit formulae for the longitudinal and transverse
relaxation times.

\begin{center}
{\LARGE Aknowledgement}
\end{center}

V.I.P., L.G.M. and N.J.M.H. gratefully acknowledge support from the
Department of Defense, DAAD 19-01-1-0592.

\begin{figure}[tbp]
\caption{ Schematic of "dynamical subsystem" - "heat bath" interaction. }
\label{fig:fig1}
\end{figure}

\begin{figure}[tbp]
\caption{ Schematic of the two-stage procedure for spin relaxation analyses }
\label{fig:fig2}
\end{figure}

\end{document}